# DCSim: Computing and Networking Integration based Container Scheduling Simulator for Data Centers

Jinlong Hu*, Zhizhe Rao, Xingchen Liu, Lihao Deng, Shoubin Dong

Guangdong Key Lab of Multimodal Big Data Intelligent Analysis, School of Computer Science and Engineering, South China University of Technology, Guangzhou, China

Email: jlhu@scut.edu.cn

Abstract

The increasing prevalence of cloud-native technologies, particularly containers, has led to the widespread adoption of containerized deployments in data centers. The advancement of deep neural network models has increased the demand for container-based distributed model training and inference, where frequent data transmission among nodes has emerged as a significant performance bottleneck. However, traditional container scheduling simulators often overlook the influence of network modeling on the efficiency of container scheduling, primarily concentrating on modeling computational resources. In this paper, we focus on a container scheduling simulator based on collaboration between computing and networking within data centers. We propose a new container scheduling simulator for data centers, named DCSim. The simulator consists of several modules: a data center module, a network simulation module, a container scheduling module, a discrete event-driven module, and a data collection and analysis module. Together, these modules provide heterogeneous computing power modeling and dynamic network simulation capabilities. We design a discrete event model using SimPy to represent various aspects of container processing, including container requests, scheduling, execution, pauses, communication, migration, and termination within data centers. Among these, lightweight virtualization technology based on Mininet is employed to construct a software-defined network. An experimental environment for container scheduling simulation was established, and functional and performance tests were conducted on the simulator to validate its scheduling simulation capabilities.

**Keywords**: Container scheduling simulation; Computing and Networking Integration; Data center

1. Introduction

The growing prevalence of cloud-native technologies, particularly containers, has led to the widespread adoption of containerized deployments within data centers. Consequently, container scheduling and management have emerged as critical technologies for the efficient operation of data centers. A significant challenge in assessing the effectiveness of container scheduling algorithms [1] in these environments is the inherent risks and costs associated with testing these algorithms in real production settings. To mitigate this issue, simulators are often utilized to assess and compare the performance of various scheduling algorithms by mimicking the operational dynamics of data centers in a controlled setting [2, 3]. By simulating diverse load scenarios and resource allocation strategies, researchers can analyze and evaluate different scheduling algorithms without disrupting the operations of the actual production system.

Most existing data center scheduling simulators concentrate on server resource allocation, load balancing, energy efficiency, and cooling system optimization [4–6]. The primary objective of these studies is to improve the overall performance and energy efficiency of data centers while minimizing operating costs. Regarding network modeling within the simulator, simplified methods are often employed, such as using predefined network parameters or disregarding network latency [7]. However, with the emergence of big data, cloud computing, and large model training and inference, networks have increasingly become a performance bottleneck for data centers. In these applications, substantial amounts of data must be transmitted between servers or to external environments. Consequently, network bandwidth, latency, and reliability significantly impact overall performance [8–10]. Traditional scheduling algorithms typically consider only the computational power constraints of containers and hosts, failing to effectively allocate resources based on the characteristics of network-intensive applications [11]. This oversight results in uneven utilization of network resources and low computational and communication efficiency. An increasing number of researchers are dedicated to developing container scheduling algorithms that enhance collaboration between computing and networking. In this context, the design of simulators capable of modeling real network environments has become essential. Professional network simulation tools such as Mininet [12], NS2 [13], and NS3 [14] can simulate complex network environments, offering precise configuration of network parameters, performance testing, and traffic analysis capabilities. By integrating these professional tools into data center simulators, the authenticity and accuracy of network characteristic simulations can be significantly improved.

In this paper, we propose a new container scheduling simulator for data centers, named DCSim. This simulator leverages the network simulation tool Mininet and the discrete event simulation library SimPy [15]. DCSim utilizes a discrete event-driven model based on SimPy to represent various discrete events, including container requests, scheduling, execution, communication, migration, and completion within a data center environment. Furthermore, DCSim integrates a network simulation module based on Mininet, which continuously monitors latency between network nodes, simulates network traffic transmission, and addresses the limitations of existing simulators in detailed network modeling.

DCSim serves as a comprehensive platform for container scheduling within data centers. It incorporates several key modules: a data center module, a network simulation module, a container scheduling module, a discrete event-driven module, and a data collection and analysis module. The platform features a three-tier application model that includes job task containers and integrates the characteristics of heterogeneous computing power into the hosts. Furthermore, DCSim offers a flexible and scalable interface for scheduling algorithms, enabling users to develop customized container scheduling algorithms tailored to various optimization objectives. DCSim models hosts and workloads within the data center modules, utilizing datasets such as Alibaba's open-source GPU workload dataset [16]. Additionally, it collects a variety of performance metrics throughout the simulation process to generate analytical reports, allowing users to evaluate the effectiveness of the scheduling algorithms.

To validate the system and facilitate user simulation experiments, DCSim implements a variety of heuristic container scheduling algorithms, including overload migration, first adaptation, loop scheduling, performance-first scheduling, and job group scheduling. Additionally, we present functional and performance tests conducted on DCSim, which demonstrate the roles of various functional modules within the simulator and their collaborative operation through experimental results.

The primary contributions are summarized as follows:
- Support for modeling heterogeneous computing power.
- Support for packet-level network modeling of hosts and containers.

- Support for custom data center hosts, workloads, and network environment configurations.
- Support discrete event modeling, which includes container requests, container scheduling, container execution, container communication, container migration, and container completion.
- Support for easy extensibility of container scheduling algorithms.

The subsequent sections of this paper are organized as follows: Section 2 provides a review of relevant simulators and conducts a comparative analysis with DCSim. Section 3 details the architecture of DCSim, clarifies its operational workflow, and elaborates on the design and implementation of each functional module within the system. Section 4 evaluates the performance of DCSim, emphasizing its effectiveness. Finally, Section 5 summarizes the content of this paper.

2. Related work

In recent years, the integration of computing and networking has become a significant focus in data centers. This integration is often referred to as Computing-aware Network (CAN), Computing First Network or Computing First Networking or Computing Force Network (CFN), or Computing Power Network (CPN) [17, 18]. Regardless of the terminology used, these concepts all tackle the challenge of optimizing the utilization of computing power resources and require an effective scheduling algorithm for these resources. This algorithm involves two fundamental dimensions: computing and networking.

Simulators lower the costs associated with theoretical research by enabling repeatable experiments in a controlled environment, while also mitigating the unpredictable risks associated with direct testing in real production settings [19]. Some simulators have been developed for scheduling. Table 1 shows the characteristics of simulators based on their computing power modeling methods, network modeling techniques, and programming languages.

CloudSim [20] is one of the most widely used data center simulators, offering modeling capabilities for virtual machines, workloads, and scheduling strategies. Built on SimJava [21], CloudSim simulates discrete events and can efficiently support large-scale data center simulations. But its topology-based network modeling [22] does not facilitate dynamic network simulation, making it challenging to effectively support simulation scenarios that depend on dynamic network environment modeling.

ContainerCloudSim [3] extends and implements Container as a Service (CaaS) based on CloudSim [23]. It adds support for container allocation, scheduling, and resource configuration, offering more precise resource isolation than virtual machines. It does not provide a detailed modeling description of container communication.

NetworkCloudSim [24] is an extension of CloudSim that emphasizes network functionality and models networks using a flow model approach. It supports applications with communication elements, bandwidth sharing, and latency, while offering the advantage of low computational overhead.

GreenCloud [25] implements packet-level network modeling using the network simulation tool NS2, enabling the implementation of a comprehensive TCP/IP protocol reference model. GreenCloud primarily focuses on researching power management solutions to create efficient and energy-saving data centers. It offers detailed modeling of energy consumption in network switches and links.

iCanCloud[26] is developed to predict the trade-off between cost and performance when executing a specific set of applications on designated hardware. iCanCloud is built upon the Simcan framework and features a graphical user interface that enables users to easily configure various data center scenarios, conduct simulation experiments, and generate graphical reports. Additionally, iCanCloud models network communication between machines and supports parallel simulation, facilitating the execution of experiments across multiple machines.

TeachCloud [27] extends the MapReduce application model and features an integrated, comprehensive workload generator called Rain, based on CloudSim. This platform allows experimentation with various cloud components, including processing elements, storage, networks, and data centers. TeachCloud supports a graphical interface that enables users to build and implement customized network topologies. Additionally, TeachCloud offers various cloud network models, such as VL2, BCube, Portland, and Dcell, to analyze the impact of different network configurations on overall system performance.

COSCO [28] and CloudSimPy [29] are developed in Python. By integrating with deep learning frameworks such as TensorFlow [30] and PyTorch [31], these tools facilitate the study of resource scheduling methods that utilize machine learning or deep learning techniques. CloudSimPy is built on the discrete event simulation framework SimPy, which offers robust event scheduling capabilities and is particularly well-suited for simulating systems with complex logic and multiple concurrent processes. But CloudSimPy does not model heterogeneous computing power or network environments. COSCO has simulated fog computing [32] scenarios and modeled container examples, but it has not provided network and discrete event modeling capabilities.

Table 1 Characteristics of different simulation tools.

| Simulator | Computing | Network | Language |
|---|---|---|---|
| CloudSim | MIPS, memory, storage, bandwidth | Topology based network modeling | Java |
| ContainerCloudSim | MIPS, memory, storage, bandwidth | Statically defined container network bandwidth | Java |
| NetworkCloudSim | MIPS, memory, storage, bandwidth | Network modeling based on flow model | Java |
| GreenCloud | MIPS, memory, storage, bandwidth | Packet level network modeling based on NS2 | C++、TCL |
| iCanCloud | MIPS, memory, storage | Packet level network modeling based on INET | C++ |
| TeachCloud | MIPS, memory, storage, bandwidth | Topology based network modeling | Java |
| COSCO | MIPS, memory, storage, bandwidth | Calculating transmission delay based on static bandwidth | Python |
| CloudSimPy | CPU core, memory | No | Python |
| DCSim | MIPS, memory, bandwidth, and FLOPS (GPU) | Packet level network modeling based on Mininet | Python |

Despite significant advancements in simulator research, several limitations remain in the area of container scheduling simulations for data centers, particularly concerning the integration of computing and networking. As the performance of computing devices increases and the volume of communication data between containers grows, network bandwidth and latency have become critical factors that cannot be overlooked during scheduling [33, 34]. However, the dynamic nature of real network environments means that traditional topologies and static definitions in network modeling fail to capture these changes. This necessitates simulators that provide packet-level network simulations to more accurately assess the performance of scheduling algorithms for computing network collaboration. This paper utilizes the network simulation tool Mininet and the discrete event simulation framework SimPy to design and implement a container scheduling simulator, DCSim, for computing and networking collaboration in data centers.

3. Design and Implementation of DCSim

3.1 System Architecture

DCSim is developed using Python, in conjunction with the network simulation tool Mininet and the discrete event simulation library SimPy. It comprises five modules: data center, network simulation, container scheduling, discrete event driving, and data collection and analysis. Figure 1 illustrates the architecture diagram of the simulator.

The data center module is responsible for reading configuration files, generating host lists, and managing workloads. The workload component generates task flows and adds newly arrived container requests to the waiting queue while maintaining both the run queue and the completion queue. The hosts within the data center are categorized into CPU servers and GPU servers based on the type of resources they provide. Each server maintains a container queue that represents all containers currently running on that server.

The network simulation module is responsible for modeling the network component of the data center. This includes creating the network topology, monitoring inter-host latency, and simulating network traffic transmission. Each data center host and its deployed containers correspond to a network node within the topology. When container communication and migration occur, the associated network traffic transmission is simulated through the relevant network nodes. The dynamic network environment is represented by changes in network latency and real-time traffic transmission rates.

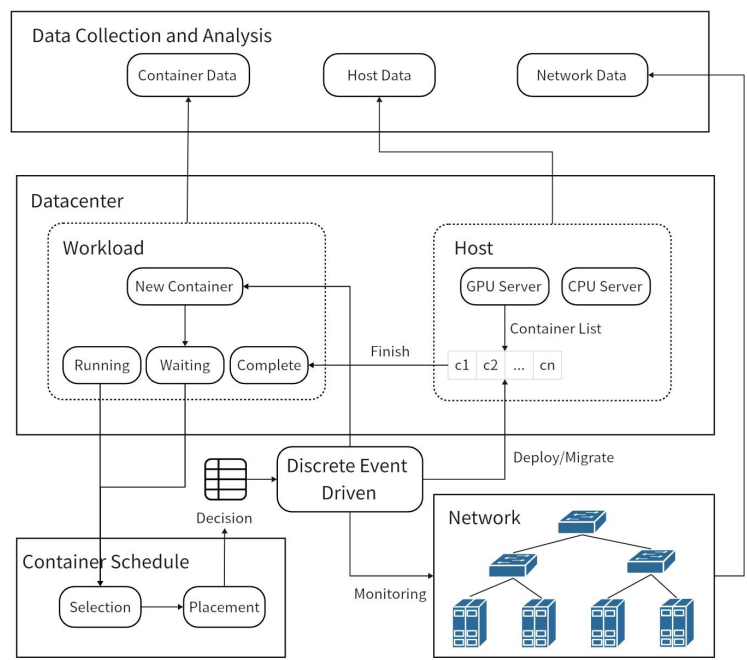

Fig. 1. Simulator architecture diagram.

The container scheduling module selects containers from the waiting queue for scheduling based on specific selection strategies or migrates containers from the running queue. The selected containers receive the final scheduling decision through placement strategies, and the simulator deploys or migrates the containers to the target host based on this decision.

The discrete event-driven module is responsible for modeling the dependencies between discrete events in the system, controlling the order of event execution, and managing the start and end of the simulation.

The data collection and analysis module is responsible for gathering host data, container data, and network data throughout the simulation process. It generates visual analysis reports at the end of the simulation to assist researchers in evaluating the performance of scheduling algorithms.

3.2  System workflow

The workflow of the system is as follows:

(1) Users configure the simulation environment, which includes performance specifications for data center hosts in CSV format, workload task flows, simulator parameters set through INI format, and custom network topologies created using Python scripts.

(2) The discrete event-driven module control simulation has commenced.

(3) The workload submodule continuously generates new container requests, which are subsequently transferred to the container scheduling module for allocation.

(4) The network simulation module continuously monitors the network latency between hosts.

(5) The data collection and analysis module continuously gathers various indicator data throughout the simulation process and records the simulation log information.

(6) When all containers finish running, stop container scheduling and network latency monitoring. The simulation process concludes, and the data collection and analysis module generates an analysis report, which is then saved to a local file.

3.3  Data Center Module

To effectively model the host and workload within the data center module, we introduce several classes, including Host, Job, Task, Container, and Workload. The Host class is designed to represent the data center's host, while a three-tier application model consisting of Job, Task, and Container is employed to simulate the job requests submitted by users. The Workload class generates container requests based on user-configured datasets and manages container queues in various states.

To model the dependency relationships between containers, we present a three-level application model consisting of job task containers, which correspond to the Job, Task, and Container classes, respectively. Users submit requests on a job-by-job basis and record both the submission and completion times. Each job corresponds to one or more tasks, and each task specifies the required amounts of CPU, memory, and GPU resources through the resource_request attribute. This attribute is a Python list containing three elements, and each task can run multiple instances. The number of instances of the task is determined by the instance_num attribute, with each instance represented as a container. The scheduler manages the scheduling on a container basis.

We categorize containers into three types: CPU-intensive, memory-intensive, and GPU-intensive, based on their distinct resource requirements. The `container_type` attribute specifies the primary resource type for each container. The `duration` property of the Container class indicates the container's runtime in seconds, while the `run_at` property tracks the elapsed execution time of the container. When `run_at` equals `duration`, the container's runtime concludes. The magnitude of each increment in `run_at` is determined by the `container_type` and the processing speed of the host. For example, if a CPU-intensive container operates on a host with a CPU speed of 2 GHz, `run_at` increases by 2 per second.

The running time of a container comprises two key components: instruction execution time and network communication time. Instruction execution time is influenced by the computational performance of the host. Deploying the container on a high-performance host can significantly reduce instruction execution time, whereas deploying it on a low-performance host will increase this time. Conversely, network communication time is affected by the network environment. When network congestion occurs or when containers communicate across multiple switches and links, communication time can increase substantially. The objective of network collaborative scheduling is to identify the optimal target host that can simultaneously minimize both instruction execution time and network communication time, thereby reducing the total running time of the container.

The status attribute of the Container class describes the state of the container, which can be one of six possible states: submit, run, communicate, migrate, wait, and complete. Table 2 outlines the distinctions among these various states of the container.

Table 2 Description of different states of containers.

| Status | Queue | Description |
| --- | --- | --- |
| inactive | undeployed | Container request submitted but not scheduled to run |
| running | deployed | The container is scheduled and deployed to run on the host |
| communicating | deployed | Trigger communication events |
| migrating | deployed, undeployed | Selected by scheduling algorithm and migrated to other hosts |
| waiting | undeployed | Suspend container operation due to migration or communication failure |
| completed | completed | End of container operation |

3.4 Network simulation module

Users create Mininet objects using custom network topology scripts to build a data center network. The network nodes in Mininet include Host, Switch, Controller, and other classes, all of which inherit from the Node class. A Host is created through the addHost interface, allowing for the specification of an IP address. It represents a virtual host, essentially functioning as a shell process that can execute Linux terminal commands via the cmd interface and monitor the execution results. Consequently, this article monitors the latency between network nodes by executing ping commands on the Host node and simulates network traffic transmission resulting from communication between containers by executing iperf commands. The schematic diagram of the network modeling structure is presented in Fig. 2.

Switch is created using the addSwitch interface for forwarding packets, while a Controller is established through the addController interface to manage the forwarding rules of packets. The integration of these two components enables Software-Defined Networking (SDN), which separates the data plane from the control plane, thereby enhancing the convenience of network management. By utilizing the addLink interface, a link can be established between a Switch and a Host, or between two Switches. Additionally, network parameters such as bandwidth, latency, and packet loss rate can be specified to simulate various network environments.

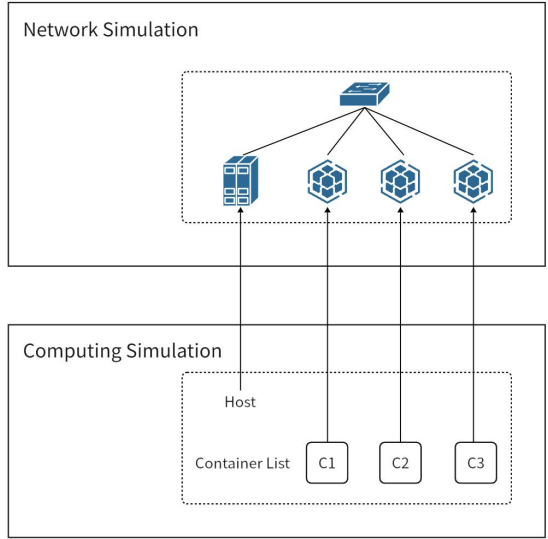

Fig. 2. Schematic diagram of network modeling structure.

The Network class receives Mininet objects created by custom network topology scripts and maintains a matrix called delay_matrix that stores the network latency between all Host nodes, as shown in Equation (1). In this context, $D_{ij}$ represents the delay_matrix and indicates the network latency between host i and host j. The Network class initializes the delay_matrix before the simulation begins and periodically updates it to ensure real-time performance. Users can configure the update interval of the delay_matrix in the configuration file.

$$D = \begin{bmatrix} D_{11} & \cdots & D_{1n} \\ \vdots & \ddots & \vdots \\ D_{n1} & \cdots & D_{nn} \end{bmatrix} \tag{1}$$

Each network node will initiate an iperf server during network initialization and run it in the background to facilitate seamless traffic transmission to that node at any time. Given that traffic transmission may fail due to network node failures, we have established a maximum number of retransmissions for traffic transmission, which users can configure. If traffic transmission fails even after exceeding the maximum number of retransmissions, it will be classified as a failure, and the container scheduling module will manage the failed traffic transmission in the future.

3.5 Container scheduling module

To model the selection, placement, and execution processes in container scheduling, the module designed in this article includes a Selection interface, a Placement interface, and an Execution interface. The Selection interface selects containers in both paused and running states, while the Placement interface selects specific hosts for these containers and collects decision-related information. The Execution interface implements placement and migration based on the decision information, which is automatically executed by the system.

For high-performance hosts selected under computing power constraints, the network environment may experience congestion, leading to increased communication times for the containers. This delay hinders the containers from minimizing their total running time while utilizing the high-performance hosts. Similarly, hosts chosen based on network constraints may fail to adequately match the required computing power. Therefore, when designing scheduling algorithms, users should consider resource constraints, host performance, primary resource types of containers, network latency, and communication rates. We present five fundamental scheduling algorithms: OverloadMigrate, First Fit, Round, PerformanceFirst, and JobGroup, to assist users in designing and conducting comparative experiments on scheduling algorithms. The five container scheduling algorithms are described as follows:

(1) OverloadMigrate: Dynamic and Resource-Aware Placement Scheme (DRAPS) [35] is a strategy that allocates containers based on the currently available resources in heterogeneous clusters and the dynamic requirements of service containers. When a specific type of resource becomes a bottleneck for a host, the system migrates the resource-intensive container to other nodes. OverloadMigrate is designed to select the container that consumes the most critical bottleneck resources for migration from overloaded hosts.

(2) FirstFit [36]: This algorithm starts with the first host in the host list and selects the first host that meets the resource constraints as the target host for scheduling the container.

(3) Round [37], the selection process begins with the next host following the previously scheduled host. The first host that meets the resource constraints will be selected as the target host for the container to be scheduled.

(4) PerformanceFirst: Based on the scheduling scheme of DRAPS [35], and taking into account the performance of host resources as well as the primary resource types of the container, PerformanceFirst is designed to select the host that provides the highest performance for the container's main resources. This selection process aims to ensure that resource constraints are satisfied while accurately identifying the target host.

(5) JobGroup: Communication Aware Worst Fit Decreasing (CA-WFD) [38] is an extension of the classic worst-fit incremental packing algorithm, specifically designed for container scheduling. Building on the CA-WFD and the container communication model introduced in this paper, JobGroup aims to select the host with the highest number of dependent containers as the target host for the scheduled containers. In cases where no dependent containers are deployed, the host with the most available resources is selected.

3.6    Discrete Event Driver Module

In order to model various discrete events and their dependencies within simulators, as well as to control the start, operation, and conclusion of the simulation, we have implemented a discrete event-driven module based on the SimPy discrete event simulation library.

This module consists of four key components: Environment, Process, Event, and Store.

SimPy's Environment is a simulation framework that is responsible for scheduling processes and events, managing the progression of simulation time, and overseeing the advancement and completion of the simulation.

The process is utilized to implement simulation models, specifically to represent the behavior of the system. These processes are standard Python generator functions that employ the `yield` keyword along with a specific event name to indicate that the process is waiting for the event to be triggered. During this time, the process will be blocked until the event occurs, at which point the process resumes execution. Table 3 outlines the processes included in this system.

An event refers to various asynchronous occurrences during the simulation process, which are used to control the operation and blocking of the process. The `SimPy.event( )` function is utilized to create an event, which can be activated by calling the `succeed()` method of the event. Each event can only be triggered once; once triggered, it will resume the process that depends on the event to a running state.

Store is a type of resource in SimPy used to model producer-consumer relationships within a system. It helps manage the dependency relationships that occur during process execution.

Table 3 Detailed description of process.

| Process | Function | Description |
| --- | --- | --- |
| Container running | run | Start when the container is deployed on the host, describing the container's runtime process |
| Container communication | communicate | Start when the container reaches the communication time, describe the container communication process |
| Container migration | migrate | Start when the container is scheduled to migrate to a new host, describe the container migration process |
| Simulation control | pre_treatment | Execute once per second to determine if the simulation has ended |
| Generate container request | generate_containers | Executed once per second to generate new container requests |
| dispatch | schedule | Execute once per second to run scheduling algorithms and deploy migration |
| Save data metrics | save_stats | Executed once per second to save indicator data |
| Update network latency | update_delay_matrix | Periodic execution, used to update the network delay matrix |

3.7 Data Collection and Analysis Module

To record event logs and metric data during the simulation process, we utilize the Stat class, which captures events occurring throughout the simulation, along with information regarding the status of hosts, containers, and networks. Various evaluation metrics are employed to assess the performance of the scheduling algorithm, culminating in the generation of an analysis report at the conclusion of the simulation. Additionally, key metric data is saved in a CSV file. We use average container response time, average container runtime, and total cost to illustrate the performance of scheduling algorithms in relation to scheduling.

4. Evaluation

4.1 System Function Testing

4.1.1 Testing Environment and Settings

The environmental testing parameters and their configurations are presented in Table 4.

Table 4 System testing environment parameters.

| Parameter fields | Parameter values |
| --- | --- |
| operating system | Ubuntu 20.04.1（64 bit） |
| CPU model | Intel Xeon E5-2673 v4 @ 2.30GHz |
| Number of CPU cores | 8 cores |
| Memory | 64GB |
| programming language | Python 3.8.10 |
| Network simulation tools | Mininet 2.3.0 |
| Discrete simulation tools | SimPy 4.0.1 |

The host information is presented in Table 5, and the workload details and simulation parameter configurations are displayed in Table 6.

Table 5 Host information.

| Category | Count | CPU Core | CPU Speed | Mem | Mem Speed | GPU | GPU Speed | Price |
|---|---|---|---|---|---|---|---|---|
| 1 | 5 | 80 | 1 | 128 | 1 | 8 | 1 | 1 |
| 2 | 5 | 80 | 2 | 128 | 2 | 8 | 2 | 1.5 |
| 3 | 5 | 80 | 3 | 128 | 3 | 8 | 3 | 3 |
| 4 | 5 | 80 | 4 | 128 | 4 | 8 | 4 | 5 |

Table 6 System test simulation configuration parameters.

| Parameter fields | Parameter values |
|---|---|
| Number of assignments | 100 |
| Number of tasks | 300 |
| Number of containers | 300 |
| Container running time | 20～30s |
| CPU request | 100～1700 (Percentage usage) |
| Memory request | 1～32 (GB) |
| GPU request | 50～200 (Percentage usage) |
| Number of communications | 1～5 |
| Data volume per communication | 100～102400 (KB) |
| Network latency matrix update cycle | 10s |
| Iperf failed retransmission count | 3 |
| Congestion determination threshold | 0.2 |
| The number of container network nodes allocated to each host | 10 |
| Overload host resource utilization threshold | 0.7 |
| Idle host resource utilization threshold | 0.3 |

We created a network topology consisting of 20 hosts based on the Spine-Leaf architecture [39], which includes 2 Spine switches, 4 Leaf switches, and 20 data center host nodes. Figure 3 illustrates the network topology utilized in the system test. The link configuration depicted in the figure is set to a default bandwidth of 1000 Mbps and a 0% packet loss rate. During the test, we adjusted various link bandwidths and packet loss rates to observe the changes in container communication time, thereby verifying the impact of different network configurations on container runtime.

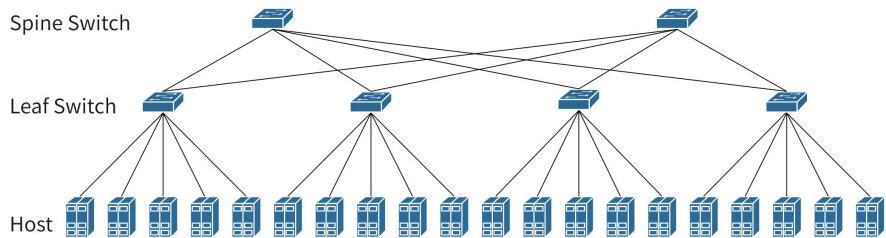

Fig. 3. System test network topology.

4.1.2   Functional module testing

The DCSim simulator consists of several key components: a data center module, a network simulation module, a container scheduling module, a discrete event-driven module, and a data collection and analysis module. The discrete event-driven module is responsible for managing the execution order of discrete events within the first three functional modules, and its effectiveness can be evaluated based on the performance of these modules. The data collection and analysis module is used to gather simulation data and generate analytical reports. Consequently, this section primarily assesses the effectiveness of the first three functional modules based on the analytical reports produced by the data collection and analysis module.

(1) Data Center Module

The data center module is responsible for generating host lists and workloads based on configuration files, as well as maintaining container queues in various states. Figure 4 illustrates the change in the number of overloaded hosts during the simulation process. From the figure, it is evident that the FirstFit scheduling algorithm, which prioritizes hosts in numerical order, causes hosts with higher numbers to enter an overloaded state earlier. Consequently, the number of overloaded hosts in the initial stage is greater than that observed with other algorithms. In contrast, the Round algorithm, due to the dispersed deployment of containers across each host, maintains a count of zero overloaded hosts from 0 to 8 seconds. As more containers complete their execution, the number of overloaded hosts for each algorithm begins to decrease around the 35-second mark.

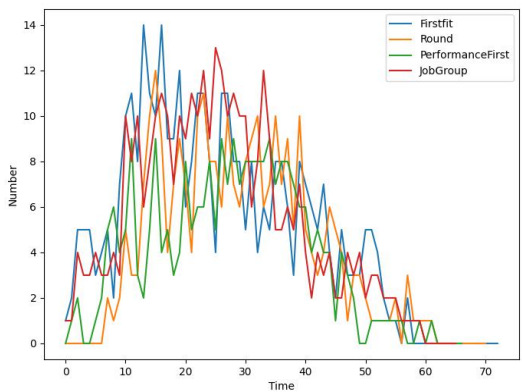
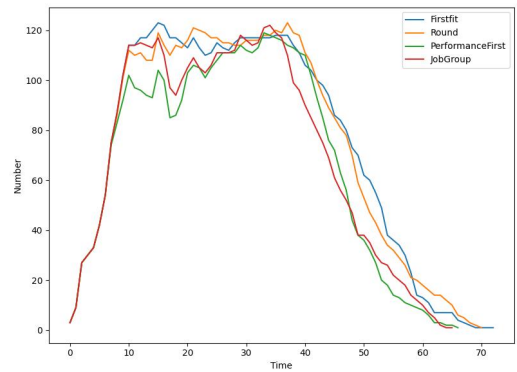

(a) Number of overloaded hosts　　　　　　　　(b) Number of containers in deployed queue

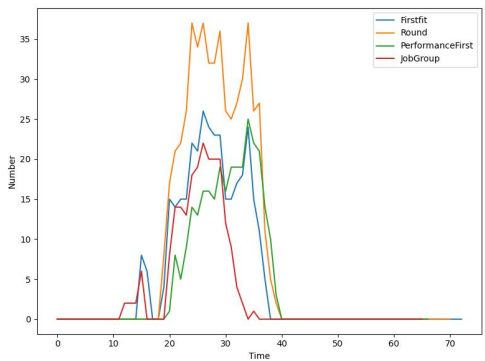
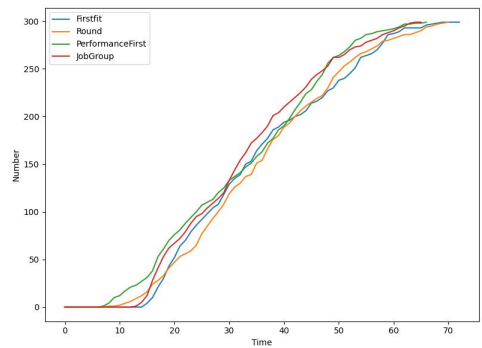

(c) Number of containers in deployed queue　　　　(d) Number of containers in deployed queue

Fig. 4. Datacenter module test results.

Figures 4(b) to 4(d) illustrate the changes in the number of containers across different container queues. The data indicates that the number of containers in the running and waiting queues for the PerformanceFirst algorithm is generally lower than that of the other algorithms, while the number of containers in the completing queue is higher. This suggests that PerformanceFirst enables containers to complete their tasks more quickly, thereby reducing the number of containers in both running and waiting states. Additionally, the number of containers in the running queue for all four scheduling algorithms began to stabilize after reaching 120, indicating that the maximum number of simultaneously running containers in the current configuration of the data center is 120.

(2) Network simulation module

The network simulation module is primarily responsible for continuously monitoring network latency between hosts and simulating data communication between containers. Figure 5 illustrates the average communication time between containers under varying link packet loss rates and bandwidths. The figure indicates that as link bandwidth decreases or link packet loss rates increase, the average communication time between containers gradually rises. Notably, the JobGroup scheduling algorithm exhibits the lowest average container communication time across different scenarios, while the Round scheduling algorithm shows the highest average container communication time. In scenarios with a link bandwidth of 200 Mbps and a packet loss rate of 2%, the differences in average container communication time among the various scheduling algorithms are most pronounced.

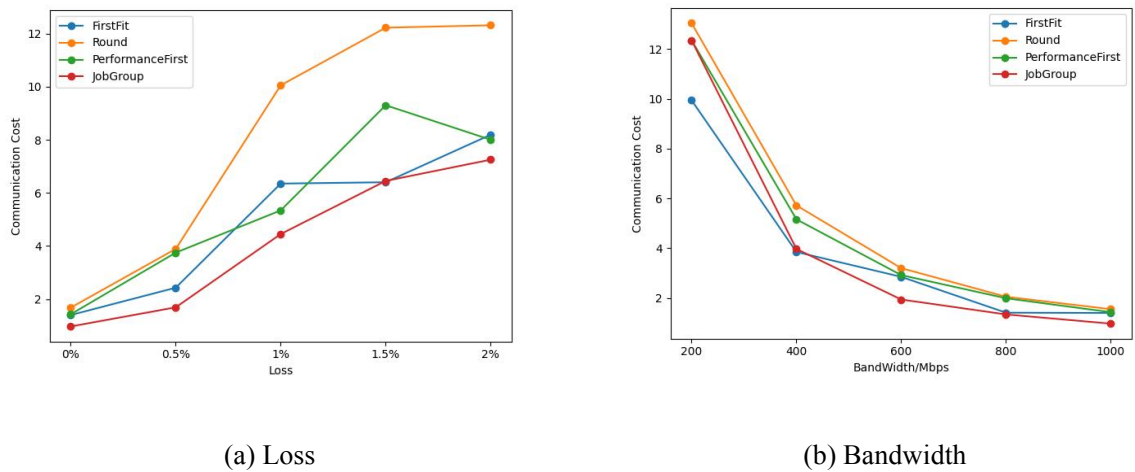

(a) Loss  (b) Bandwidth

Fig. 5. Average container communication time under different link loss or link bandwidth.

(3) Container scheduling module

The container scheduling module is primarily responsible for selecting containers to be scheduled based on various scheduling algorithms, as well as deploying or migrating containers to run on the host. Figure 6 illustrates the number of newly arrived containers in each scheduling round and the corresponding scheduling decisions made by the algorithm. The figure indicates that no new container requests are generated once the simulation reaches 39 seconds. During the initial 0-10 second period, the host's available resources are sufficient, allowing new containers to be deployed and run immediately by the different scheduling algorithms upon arrival. Consequently, the number of scheduling decisions closely matches the number of new containers. However, after 10 seconds, the performance differences among the various scheduling algorithms begin to emerge. In the 10-20 second period, containers scheduled earlier through the PerformanceFirst and JobGroup algorithms demonstrate improved performance, leading to a slight increase in the number of scheduling decisions compared to other algorithms, as host resources are released. By around 40 seconds, the number of scheduling decisions for all four algorithms drops to zero, indicating that there are no containers awaiting scheduling at that time. Reason: Improved clarity, vocabulary, and technical accuracy while maintaining the original meaning.

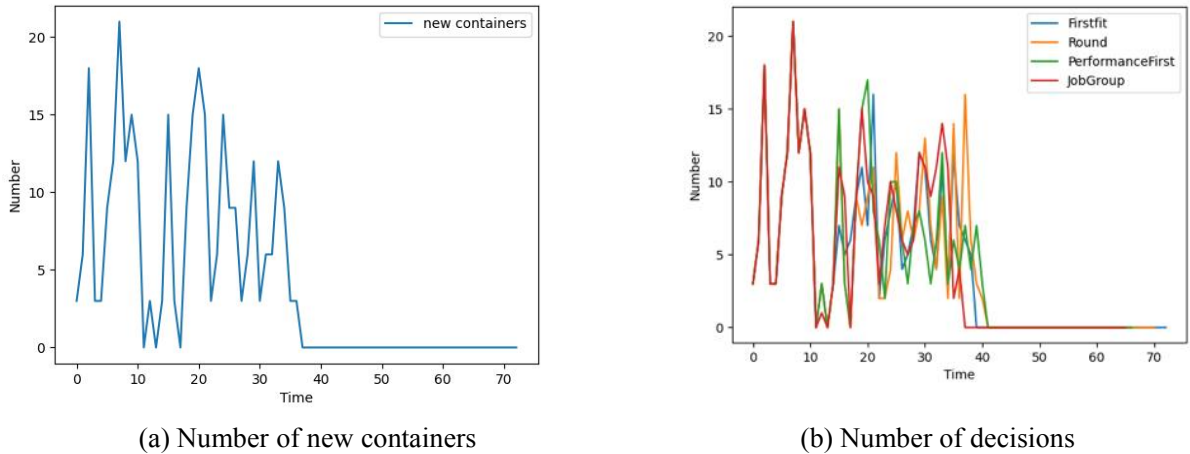

(a) Number of new containers          (b) Number of decisions

Fig. 6. Test results of container scheduling module.

Figure 7 illustrates the number of containers selected for migration in each scheduling round during the simulation experiment using OverloadMigrate. The figure indicates that during the 0-40 second period, the continuous influx of new container requests caused the data center host to quickly reach an overload state. Due to the lack of available idle hosts, the overload migration algorithm selected very few containers for migration. In the 45-55 second period, the absence of new container requests, combined with the release of host resources as more containers were executed, allowed the overload migration algorithm to begin selecting containers for migration. After 55 seconds, as the number of concurrently running containers decreased, the data center host was no longer in an overloaded state, resulting in the overload migration algorithm ceasing to select containers for migration.

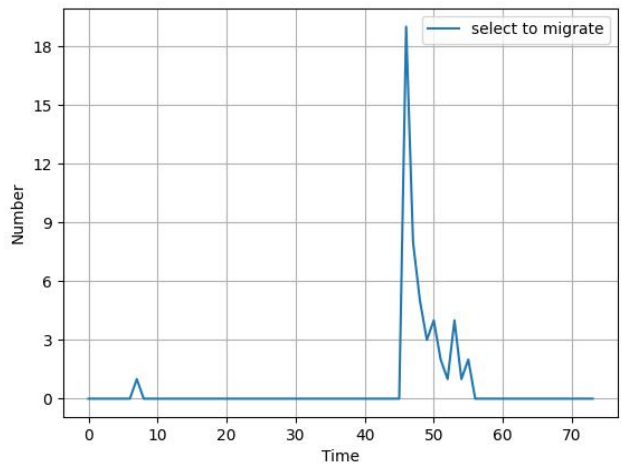

Fig. 7. OverloadMigrate test results.

### 4.1.3 Comprehensive System Evaluation

This section will conduct a comprehensive test of the simulator to verify the effectiveness of collaboration among various functional modules. By comparing the simulation evaluation indicators of different basic scheduling algorithms, we will assess the performance of these algorithms in reducing average container running time, minimizing total data center costs, and balancing loads.

Figure 8 illustrates the experimental results of the simulation under varying link packet loss rates. The data indicates that the Round algorithm, by distributing containers across different hosts, increases the communication time between containers, resulting in a higher average container running time compared to the other algorithms. Furthermore, as the packet loss rate escalates, this disparity becomes more pronounced. In contrast, the JobGroup algorithm aims to deploy containers on the same host whenever possible, thereby minimizing communication time between containers and achieving the lowest average container running time among the four algorithms. Generally, the FirstFit algorithm places contiguous containers on the same or adjacent hosts, which also helps to reduce communication time between containers to some extent. Consequently, its performance ranks just below that of JobGroup.

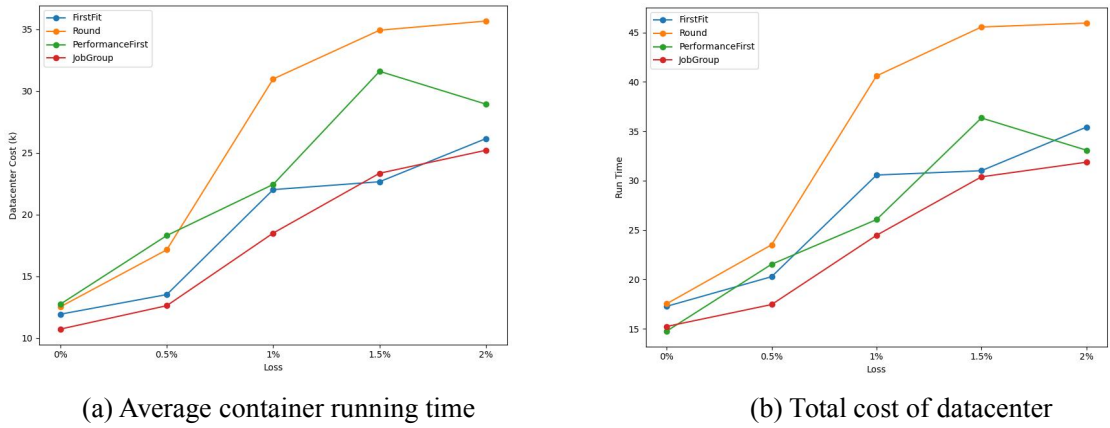

(a) Average container running time    (b) Total cost of datacenter

Fig. 8. Overall system test results.

The workload utilized in the aforementioned experiment will cause the data center host to be fully loaded for an extended period during the initial phase of the simulation. As the container operates, most hosts will remain idle in the later stages of the simulation, which complicates the assessment of the performance of various scheduling algorithms in load balancing. Consequently, this section has adjusted the job arrival rate in the workload, extending the original 100 jobs that arrived within 36 seconds to arrive over a span of 100 seconds, while maintaining all other configurations unchanged for the experiment.

Figure 9 illustrates the variation in the number of containers across different queues. It is evident from the figure that, compared to the period before the workload modification, the number of containers in the running queue has significantly decreased, while the number of containers in the waiting queue has consistently remained at zero.

Figure 10 illustrates the variance in resource utilization within a data center under two distinct workload configurations. The figure indicates that, due to Round's dispersed deployment of containers throughout the data center, JobGroup selects the host with the highest available resources when scheduling the first container of the job. These two algorithms exhibit lower variance in resource utilization compared to the other two algorithms, with the difference being more pronounced under the workload configuration that arrives within 100 seconds.

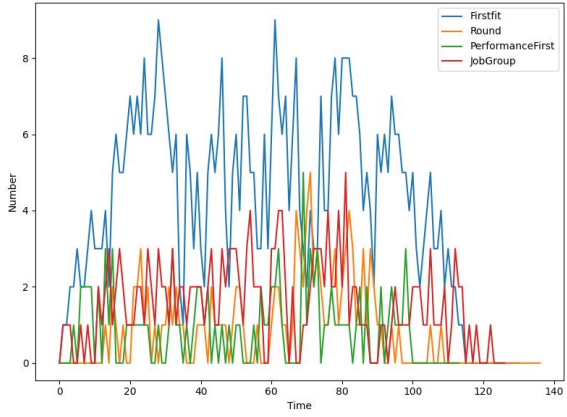 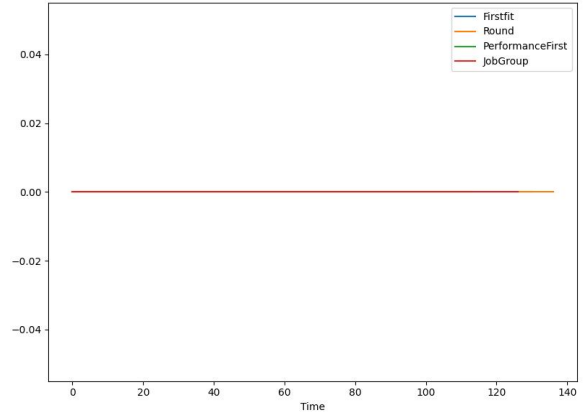

(a) deployed queue　　　　　　　　　　　　(b) undeployed queue

Fig. 9. Number of containers in different queues.

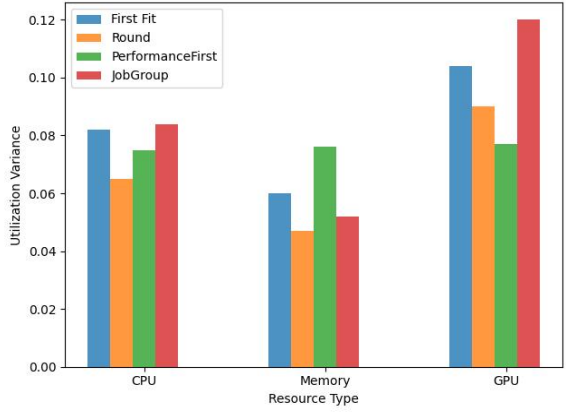 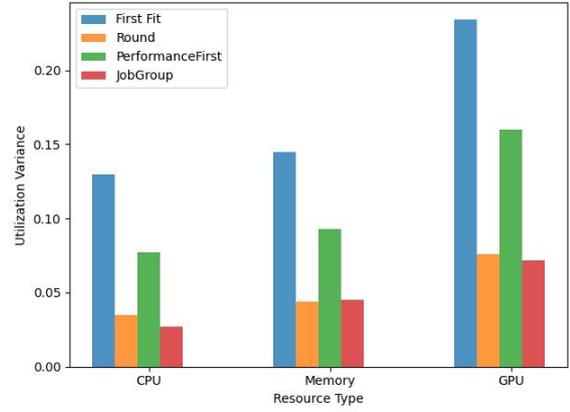

(a) 36s workload　　　　　　　　　　　　(b) 100s workload

Fig. 10. Variance of resource utilization under different workloads.

### 4.2 System performance experiment

#### 4.2.1 Experimental Environment and Settings

The performance experiment environment is consistent with the system function testing environment. It is built on the Ubuntu 20.04.1 operating system and is equipped with 8 CPU cores and 64 GB of memory.

The scheduling algorithm used in this section is the collaborative container migration algorithm utilized in the previous migration experiment. The evaluation metrics for system performance experiments primarily include simulation time, CPU utilization, and memory usage. Simulation time is further categorized into network initialization time and simulation tuning time. Network initialization time encompasses the creation of network nodes and links, as well as the initiation of the IPerf server on the host node, which is largely dependent on the number of network nodes. Conversely, simulation scheduling time refers to the duration from the completion of network initialization to the conclusion of the simulation, which is primarily influenced by the number of jobs and request times within the workload. To assess the simulator's performance across various host and workload scales, this section of the experiment employed five distinct host and workload configurations. The specific configuration details are presented in Table 7, where the number of containers corresponding to the network node counts of 200, 400, 600, 800, and 1000 are 300, 600, 900, 1200, and 1500, respectively.

Table 7 System performance experimental simulation configuration parameters.

| Parameter fields | Parameter values |
| --- | --- |
| Number of hosts | 20、40、60、80、100 |
| Number of network nodes | 200、400、600、800、1000 |
| Number of assignments | 100、200、300、400、500 |
| Number of tasks | 300、600、900、1200、1500 |
| Number of containers | 300、600、900、1200、1500 |

Due to the network simulation tool Mininet simulates a network node by creating Linux processes, a significant number of Linux processes are generated during the simulation. To collect CPU and memory usage data throughout the simulation, this section of the experiment utilizes Python's psutil library. First, it records the CPU and memory usage of the experimental machine during idle periods. Next, the simulator is executed, and the CPU and memory usage of the experimental machine are recorded every second. Finally, the collected data is subtracted from the idle time data to determine the CPU and memory usage required for the simulator to operate.

4.2.2    Experimental results and analysis

Figure 11 illustrates the simulation time, network initialization time, CPU utilization, and memory usage during the operation of the simulator with network node counts of 200, 400, 600, 800, and 1000, respectively.

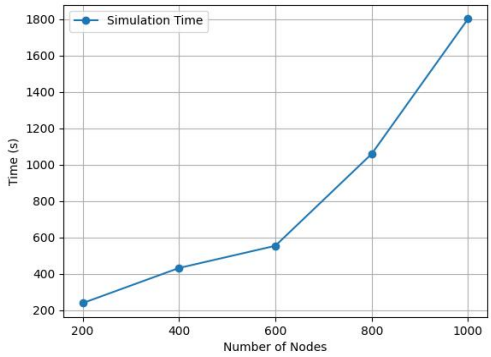

(a) Simulation time

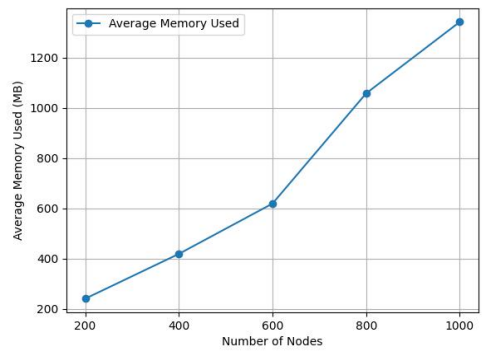

(b) Network initialization time

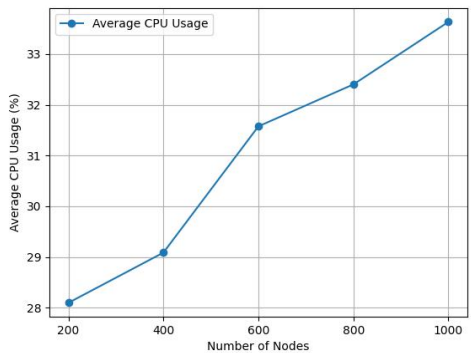

(c) CPU utilization rate

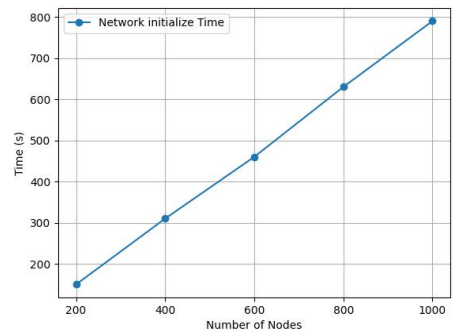

(d) Memory usage

Fig. 11. System performance experimental results.

The results indicate a linear relationship between network initialization time and the number of network nodes. Each network node creation takes approximately 0.8 seconds, and the total simulation time comprises both network initialization time and simulation scheduling time. Consequently, the simulation time is also influenced by the workload size. As the number of network nodes and the workload increases, the rate of increase in simulation time becomes more pronounced. The system's CPU utilization is maintained at around 30% and gradually rises with the addition of network nodes. When simulating 1,000 network nodes, CPU utilization increases by 19.71% compared to simulating 200 network nodes. In contrast to CPU utilization, the rise in memory usage is more substantial, as each network node operates as a separate Linux process. Therefore, as the number of simulated network nodes increases, the required memory also doubles. Simulating 1,000 network nodes consumes an average of 1,342 MB of memory, which is 488% more than the memory used for simulating 200 network nodes. Although network simulation using Mininet can provide more accurate and dynamic network modeling, it also incurs greater CPU and memory overhead.

5. Conclusions

We designed and implemented a container scheduling simulator, DCSim, for data centers, utilizing the Mininet network simulation tool and the SimPy discrete event simulation library. The proposed simulator consists of several key components: a data center module, a network simulation module, a container scheduling module, a discrete event-driven module, and a data collection and analysis module. Users can simulate various container scheduling scenarios by modifying configuration files. The system provides heterogeneous computing power modeling and dynamic network simulation capabilities, enabling continuous monitoring of changes in network latency between nodes. It simulates network traffic generated by container communication and migration, offering essential support for container scheduling algorithms in computational network collaboration. Additionally, the system implements a range of fundamental container scheduling algorithms for users to utilize in simulation experiments. It also features scalable scheduling algorithm interfaces and visual analysis reports of simulation results, allowing users to design and evaluate scheduling algorithms with diverse optimization objectives. Reason: Improved clarity, vocabulary, and technical accuracy while maintaining the original meaning.

To evaluate the proposed simulator, this paper conducted functional and performance tests, offering a comprehensive overview of the experimental environment and simulation settings. The results of the functional testing demonstrated the effectiveness of the data center module, network simulation module, container scheduling module, discrete event-driven module, and data collection and analysis module. Furthermore, the performance test results underscored the simulator's efficiency in terms of simulation time, network initialization time, CPU utilization, and memory usage.

References

[1] Zhou R, Li Z, Wu C. Scheduling frameworks for cloud container services. IEEE/acm transactions on networking, IEEE, 2018, 26(1): 436–450.

[2] Sharkh M A, Kanso A, Shami A, et al. Building a cloud on earth: A study of cloud computing data center simulators. Computer Networks, Elsevier, 2016, 108: 78–96.

[3] Piraghaj S F, Dastjerdi A V, Calheiros R N, et al. ContainerCloudSim: An environment for modeling and simulation of containers in cloud data centers. Software: Practice and Experience, 2017, 47(4): 505–521.

[4] Wickremasinghe B, Calheiros R N, Buyya R. Cloudanalyst: A cloudsim-based visual modeller for analysing cloud computing environments and applications. 2010 24th IEEE international conference on advanced information networking and applications. IEEE, 2010: 446–452.

[5] Gupta S K, Gilbert R R, Banerjee A, et al. Gdcsim: A tool for analyzing green data center design and resource management techniques. 2011 International Green Computing Conference and Workshops. IEEE, 2011: 1–8.

[6] Ostermann S, Plankensteiner K, Prodan R, et al. GroudSim: An Event-Based Simulation Framework for Computational Grids and Clouds. M.R. Guarracino, F. Vivien, J.L. Träff, et al. Euro-Par 2010 Parallel Processing Workshops. Berlin, Heidelberg: Springer Berlin Heidelberg, 2011, 6586: 305–313.

[7] Sonmez C, Ozgovde A, Ersoy C. EdgeCloudSim: An environment for performance evaluation of edge computing systems. Transactions on Emerging Telecommunications Technologies, 2018, 29(11): e3493.

[8] Mansouri Y, Prokhorenko V, Babar M A. An automated implementation of hybrid cloud for performance evaluation of distributed databases. Journal of Network and Computer Applications, 2020, 167: 102740.

[9] Li M, Andersen D G, Smola A J, et al. Communication Efficient Distributed Machine Learning with the Parameter Server. Advances in Neural Information Processing Systems. Curran Associates, Inc., 2014, 27.

[10] Li M, Andersen D G, Park J W, et al. Scaling Distributed Machine Learning with the Parameter Server. 2014: 583–598.

[11] Paščinski U, Trnkoczy J, Stankovski V, et al. QoS-Aware Orchestration of Network Intensive Software Utilities within Software Defined Data Centres: An Architecture and Implementation of a Global Cluster Manager. Journal of Grid Computing, 2018, 16(1): 85–112.

[12] GitHub - mininet/mininet: Emulator for rapid prototyping of Software Defined Networks. /2024-04-05. https://github.com/mininet/mininet.

[13] Issariyakul T, Hossain E. Introduction to Network Simulator 2 (NS2). Introduction to Network Simulator NS2. Boston, MA: Springer US, 2009: 1–18.

[14] Documentation | ns-3. /2024-04-06. https://www.nsnam.org/documentation/.

[15] Overview - SimPy 4.1.1 documentation. /2024-04-05. https://simpy.readthedocs.io/en/latest/.

[16] clusterdata/cluster-trace-gpu-v2020 at master · alibaba/clusterdata · GitHub. /2024-04-05. https://github.com/alibaba/clusterdata/tree/master/cluster-trace-gpu-v2020.

[17] Król M, Mastorakis S, Oran D, et al. Compute first networking: Distributed computing meets ICN, Proceedings of the 6th ACM Conference on Information-Centric Networking. 2019: 67-77.

[18] Gong X, Bai C, Ren S, et al. A Survey of Compute First Networking, 2023 IEEE 23rd International Conference on Communication Technology (ICCT). IEEE, 2023: 688-695

[19] Handigol N, Heller B, Jeyakumar V, et al. Reproducible network experiments using container-based emulation. Proceedings of the 8th international conference on Emerging networking experiments and technologies. Nice France: ACM, 2012: 253–264.

[20] Calheiros R N, Ranjan R, Beloglazov A, et al. CloudSim: a toolkit for modeling and simulation of cloud computing environments and evaluation of resource provisioning algorithms. Software: Practice and Experience, 2011, 41(1): 23–50.

[21] Mallet F. SimJava. School of Informatics, The University of Edinburgh/2024-04-05. https://www.icsa.inf.ed.ac.uk/research/groups/hase/simjava/.

[22] Medina A, Lakhina A, Matta I, et al. BRITE: An approach to universal topology generation. MASCOTS 2001, Proceedings Ninth International Symposium on Modeling, Analysis and Simulation of Computer and Telecommunication Systems. IEEE, 2001: 346–353.

[23] Elahi B, Malik A W, Rahman A U, et al. Toward scalable cloud data center simulation using high‐level architecture. Software: Practice and Experience, 2020, 50(6): 827–843.

[24] Garg S K, Buyya R. Networkcloudsim: Modelling parallel applications in cloud simulations. 2011 Fourth IEEE International Conference on Utility and Cloud Computing. IEEE, 2011: 105–113.

[25] Kliazovich D, Bouvry P, Khan S U. GreenCloud: a packet-level simulator of energy-aware cloud computing data centers. The Journal of Supercomputing, 2012, 62(3): 1263–1283.

[26] Núñez A, Vázquez-Poletti J L, Caminero A C, et al. iCanCloud: A Flexible and Scalable Cloud Infrastructure Simulator. Journal of Grid Computing, 2012, 10(1): 185–209.

[27] Jararweh Y, Alshara Z, Jarrah M, et al. TeachCloud: a cloud computing educational toolkit. International Journal of Cloud Computing, 2013, 2(2/3): 237.

[28] Tuli S, Poojara S R, Srirama S N, et al. COSCO: Container orchestration using co-simulation and gradient based optimization for fog computing environments. IEEE Transactions on Parallel and Distributed Systems, IEEE, 2021, 33(1): 101–116.

[29] Li F, Hu B. DeepJS: Job Scheduling Based on Deep Reinforcement Learning in Cloud Data Center. Proceedings of the 2019 4th International Conference on Big Data and Computing - ICBDC 2019. Guangzhou, China: ACM Press, 2019: 48–53.

[30] Abadi M, Barham P, Chen J, et al. TensorFlow: a system for Large-Scale machine learning. 12th USENIX symposium on operating systems design and implementation (OSDI 16). 2016: 265–283.

[31] Paszke A, Gross S, Massa F, et al. Pytorch: An imperative style, high-performance deep learning library. Advances in neural information processing systems, 2019, 32.

[32] Gupta H, Vahid Dastjerdi A, Ghosh S K, et al. iFogSim: A toolkit for modeling and simulation of resource management techniques in the Internet of Things, Edge and Fog computing environments. Software: Practice and Experience, 2017, 47(9): 1275–1296.

[33] Rodrigues L R, Pasin M, Alves O C, et al. Network-aware container scheduling in multi-tenant data center. 2019 IEEE Global Communications Conference (GLOBECOM). IEEE, 2019: 1–6.

[34] Santos J, Wang C, Wauters T, et al. Diktyo: Network-aware scheduling in container-based clouds. IEEE Transactions on Network and Service Management, IEEE, 2023.

[35] Mao Y, Oak J, Pompili A, et al. Draps: Dynamic and resource-aware placement scheme for docker containers in a heterogeneous cluster. 2017 IEEE 36th International Performance Computing and Communications Conference (IPCCC). IEEE, 2017: 1–8.

[36] Dósa G, Sgall J. First Fit bin packing: A tight analysis. DROPS-IDN/v2/document/10.4230/LIPIcs.STACS.2013.538. Schloss Dagstuhl – Leibniz-Zentrum für Informatik, 2013.

[37] Swarm mode overview. Docker Documentation. 2024-04-05. https://docs.docker.com/engine/swarm/.

[38] Lv L, Zhang Y, Li Y, et al. Communication-aware container placement and reassignment in large-scale internet data centers. IEEE Journal on Selected Areas in Communications, IEEE, 2019, 37(3): 540–555.

[39] Sultan M, Imbuido D, Patel K, et al. Designing knowledge plane to optimize leaf and spine data center. 2020 IEEE 13th International Conference on Cloud Computing (CLOUD). 2020: 13–15.